\documentclass[aps,footinbib,twocolumn,showpacs,amsmath,amssymb,superscriptaddress,prb,groupedaddress]{revtex4}
\usepackage{graphicx,amsmath}
\usepackage{epstopdf}
\usepackage{amssymb}
\usepackage{grffile}
\usepackage[usenames]{color}
\usepackage{indentfirst}
\usepackage{float}
\usepackage{color}
\usepackage{mathrsfs}
\usepackage{dcolumn}
\usepackage{bm}
\usepackage[colorlinks=true,bookmarks=false,citecolor=blue,linkcolor=red,urlcolor=blue]{hyperref}

\begin{document}
\title{The Monte Carlo simulation of the topological quantities in FQH systems}
\author{Yi Yang}
\author{Zi-Xiang Hu}
\email{zxhu@cqu.edu.cn}
\affiliation{Department of Physics and Chongqing Key Laboratory for Strongly Coupled Physics, Chongqing University, Chongqing 401331, People's Republic of China}


\begin{abstract}
Generally speaking, for a fractional quantum Hall (FQH) state, the electronic occupation number for each Landau orbit could be obtained from numerical methods such as exact diagonalization, density matrix renormalization group or algebraic recursive schemes (Jack polynomial). In this work, we apply a  Metroplis Monte Carlo method to calculate the occupation numbers of several FQH states in cylinder geometry. The convergent occupation numbers for more than 40 particles are used to verify the chiral bosonic edge theory and determine the topological quantities via momentum polarization or dipole moment. The guiding center spin, central charge and topological spin of different topological sectors are consistent with theoretical values and other numerical studies. Especially, we obtain the topological spin of $e/4$ quasihole in Moore-Read and 331 states. At last, we calculate the electron edge Green's functions and analysis position dependence of the non-Fermi liquid behavior. 
\end{abstract}
\date{\today } 
\maketitle

\section{Introduction}
Since the discovery of fractional quantum Hall effect (FQHE), its rich physical connotations and novel topological properties have attracted the extensive attention of physicists.~\cite{DCTHLSACG,RBL} Different from integer quantum Hall state (IQHE), FQH state is embedded with quantum topological order which manifests novel properties include fractional charge excitation, fractional statistics, topological ground state degeneracy, gapless chiral edge excitation and topological entanglement entropy, etc. ~\cite{XGWQN,WXG3,DAJRSFW,FDMHEHR} The quantum Hall bulk is an incompressible insulator which is difficult to produce a signal in experimental measurements. Therefore, its conducting gapless edge mode provides a window to detect the topological properties due to the mechanism of bulk-edge correspondence.~\cite{XGW,XGWQN} Early in 1990s, the physics of edge excitation was considered extremely important for studying the FQHE.~\cite{AHMD,WXG1,WXG2,WXG3} It is known that most of the FQH edges can be treated as a chiral Luttinger liquid ($\chi LL$) instead of a non-interacting Fermi-liquid.~\cite{MHDCTMGLNPKWW}  In experiments, one can measure the non-Fermi liquid behavior via nonlinear $I \propto V^{\alpha}$ relation in the tunneling experiment from Fermi-liquid to FQH liquid.  The Tomonage-Luttinger (TL) exponent $\alpha$ could be calculated from the edge Green's function $G(|\vec{r}_1- \vec{r}_2|) = \langle \psi^\dagger(\vec{r}_1)\psi(\vec{r}_2)\rangle \propto |\vec{r}_1- \vec{r}_2|^{-\alpha}$. The edge electron propagator also can describe the entanglement of two particles on the edge. Wen's effective theory~\cite{WXG3} demonstrated that the spatial decay of electron propagator involves a non-Fermi-liquid exponent $\alpha=q$ for $\nu = 1/q$ Laughlin state, $\alpha = 3$ for MR state and $331$ state. For a  realistic system with Coulomb interaction, the values of $\alpha$ are found not that universal. This has attracted a lot of theoretical and experimental attentions,~\cite{MGDCTLNPKWWAMC,AMCMKWCCCLNPKWW,AMCLNPKWW,AMC,MG, VJGEVT,SSMJKJ1,XWFEEHR,Chamon94,XWKYEHR,KY,YNJHKNGM,MHDCTMGLNPKWW} such as the influence of the edge reconstruction, the sample qualities and the emergence of neutral mode. Recently, it was verified that the FQH in suspended graphene could avoid those obstructive factors and realize the universal edge physics.~\cite{ZXHRNBXWKY, Kumar21, Kumar22a, Kumar22b} Similarly, the occupation numbers near the edge obey $\displaystyle\lim_{k \rightarrow edge}n_{k} \propto k^{\beta}$ in continuum limit, as predicted by chiral boson edge theory.~\cite{WXG4} At the same time, the information of the bulk magnetoroton excitation has been claimed to be embodied in the oscillation of the occupation numbers near the edge.~\cite{Shibata}

 In a correlated FQH system, the density deviates  from the bulk filling $\rho = \frac{\nu}{2\pi l_B^2}$ near the edge and thus results in an extra ``intrinsic dipole moment" which is related to the guiding-center Hall viscosity.~\cite{JEARSPGZ,Park,F. D. M. Haldane} It is worth mentioning that the Hall viscosity is characterized by a rational number and a metric tensor that defines distances on an ``incompressibility length-scale", and its magnitude provides a lower bound to the coefficient of the $O(q^4)$ small-$q$ limit of the guiding center structure factor. The Hall viscosity is also related to the momentum polarization~\cite{Read09, HHTYZXLQ}  of the system while rotating half of the system and keeping another half invariant. In fact, the momentum polarization is the sub-leading term of the average value of a ``partial translation operator". Therefore, the calculation of the intrinsic dipole moment, or the momentum polarization, is averaging the momentum operator of a subsystem in a bipartition. The interesting thing is that the topological quantities of the FQH state, such as the guiding-center spin, central charge and the topological spin of the quasiparticle excitation could also be determined from the coefficients and corrections of the momentum polarization.~\cite{MPZRSKMFP, Park,LDHWZ, LHZLDNSFDMHWZ} The guiding-center spin is related to the non-dissipative response of the metric perturbation in FQH liquids. Its coupling with the geometric curvature of the underlying manifold gives the topological shift of the FQH states in spherical geometry. Then topological spin and central charge are the elements of modular-$T$ matrix which are used to describe the topological order of the FQH state.~\cite{EKVXGW} Meanwhile, the central charge determines the heat current $I_E=\frac{\pi}{6}cT^2$ at a given temperature~\cite{IA} and is also related to the gravitational anomaly of the edge.~\cite{LAGEW}

In numerical calculation, the density fluctuations of the quantum Hall edge affect several Landau orbitals and the scope of its influence becomes larger for small bulk density. For example, the edge of $\nu = 1/5$ Laughlin state affects more orbits than that of the $\nu = 1/3$ Laughlin state. Moreover, in case of the realistic long range Coulomb interaction, the edge oscillates deeper into the bulk than the short range model interaction. A criteria to get a full profile of the edge state is the density in the bulk should be stable at the filling factor $\nu/2\pi l_B^2$. This is usually out of the reach of exact diagonalization or the Jack polynomial which are limited by the small size of the Hilbert space.
The inaccurate momentum polarization calculation in small system size can not give the convergence of the physical quantities and even wrong results at some time. The developments to solve this problem have been done are the density-matrix renormalization group~\cite{Shibata,LHZLDNSFDMHWZ} and Matrix product state description~\cite{MPZRSKMFP}. In this work, we
develop the Monte Carlo simulation method in cylinder geometry to calculate the occupation numbers for several FQH states. From these occupation numbers, we explore the momentum polarization and its related topological quantities in a high accuracy. The edge Green's function is also calculated for large system and the parameter of the chiral Lutteringer liquid theory is determined in a higher accuracy than the previous studies.

The rest of the paper is organized as follows. In Sec.~\ref{sec2}, we calculate the occupation numbers for several FQH states in cylinder geometry and revisit the exponents of the chiral boson theory (CBT) of the edge.  In Sec.~\ref{sec3}, we calculate the topological quantities from edge dipole moment and momentum polarization.  In Sec.~\ref{sec4}, we acquire the TL exponent $\alpha$ from equal time Green's function and discuss the validity of the TL theory. The MR and 331 states are also considered. Sec.~\ref{sec:discussion} gives the conclusions and discussions.

\section{The occupation number and its scaling behaviours}
\label{sec2}

\begin{figure}[ht]        
\center{\includegraphics[width=8.5cm]  {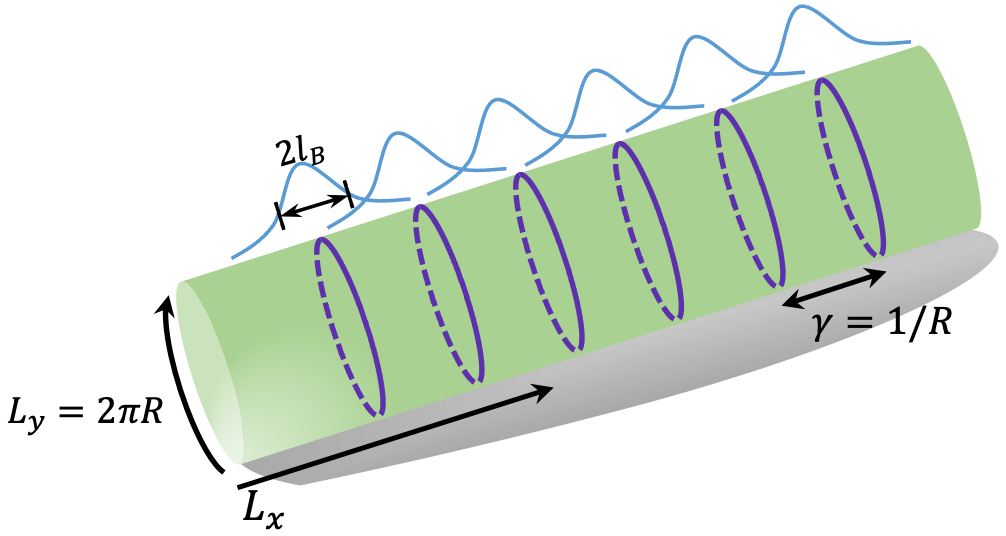}}        
\caption{\label{cylindermodel} The sketch map of cylinder model with $L_x$ and $L_y$ in two directions. $L_y=2\pi R$ is the circumference of cylinder, and the reciprocal of the radius is defined as $\gamma = 1 / R$.} 
\end{figure}

We firstly introduce the occupation number calculation by Metropolis Monte Carlo which was previously implemented in disk geometry.~\cite{Morf, SMAHMD,UKMGYG} The disk geometry in symmetric gauge has unequally spaced orbits which evolves much more orbitals for edge profile and induces to slow the convergence of the bulk density. In this work, as shown in the sketch of Fig.~\ref{cylindermodel}, we use the cylinder geometry which has advantages that the space between adjacent Landau orbits is homogeneous and the length of the edge on the two ends is tunable by varying the aspect ratio $L_x / L_y$ with keeping the surface area invariant.  The normalized $N$-electron Laughlin wave function $|\psi^c_{1/q} \rangle$ at filling $\nu = 1/q$ is~\cite{SJEHLRS,D. J Thouless}
\begin{eqnarray}
 \label{eq:a1}
    && |\psi^c_{1/q} \rangle  = \frac{1}{\sqrt{N!}}\frac{1}{(2\pi \gamma^{-1} \sqrt{\pi})^{N/2}} \exp(-\frac{9}{2}\gamma^2 \sum_{j=0}^{N-1}j^2 )  \nonumber \\
      &&\prod_{j<k} (e^{\gamma \mathbf{z}_j}-e^{\gamma \mathbf{z}_k})^q e^{-\frac{1}{2}\sum_{i=1}^N x_i^2} e^{-\sum_i \frac{q\gamma }{2}(N-1)\mathbf{z_i}} 
\end{eqnarray}
in which $\mathbf{z}_i=(x_i+iy_i)/l_B$ is the coordinate of the $i$'th
 particle, $l_B$ is the magnetic length $l_B = \sqrt{\hbar/eB}$ which we set to one and the Landau orbial space $\gamma = 2\pi/L_y = 1/R$ where $R$ is the radius of the cylinder. The last term is a global shift which lets the FQH state be symmetric around the center of cylinder at $x=0$. The average occupation of the $m$th single-paricle state is
\begin{equation}
	\begin{split}
		\langle c_m^{\dagger}c_m \rangle_{1/q}&=\frac{\langle \psi^c_{1/q} |c_m^{\dagger}c_m|\psi^c_{1/q} \rangle  }{\langle \psi^c_{1/q}|\psi^c_{1/q} \rangle} \\
		&=\int d^2 z_1 d^2z_2 \rho_{1/q}(\mathbf{z_1},\mathbf{z_2})\phi_m^{c*}(\mathbf{z_1}) \phi^c_m(\mathbf{z_2} )
	\end{split}
\end{equation}
where $\rho_{1/q}$ is the one-particle density matrix~\cite{Jain} and $\phi^c_m(\mathbf{z})=\frac{1}{\sqrt{\pi^{1/2}L_y}}e^{i ky}e^{-(x-k)^2/2}$ is the wave package of the lowest Landau level in Landau gauge with wave vector $k=\frac{2\pi m }{L_y}$. Now the $y$ direction translation momentum quantum number $m$'s are symmetrically distributed in range $[-\frac{q(N-1)}{2}, \frac{q(N-1)}{2}]$. The one-particle density matrix is
\begin{eqnarray} 
  \label{eq:a3}
	 \rho_{1/q}(\mathbf{z_a},\mathbf{z_b}) &=& N\int d^2 \mathbf{z_2} \cdots \int d^2 \mathbf{z_N} \psi^c_{1/q}(\mathbf{z_a,z_2\cdots z_{N}}) \nonumber \\
	&&\psi_{1/q}^{c*}(\mathbf{z_b,z_2\cdots z_{N}})/\int \prod_{i=1}^{N} d^2\mathbf{z_i}|\psi^c_{1/q}|^2.
\end{eqnarray} 
Because $\phi^c_m$ and $|\psi^c_{1/q} \rangle$ conserve the translation momentum operator along $y$, the one-particle density matrix could be written in second quantized form
\begin{equation}
	\rho_{1/q}(\mathbf{z_a},\mathbf{z_b})=\sum_m \langle c_m^{\dagger}c_m \rangle_{1/q} \phi^c_m(\mathbf{z_a}) \phi_m^{c*}(\mathbf{z_b})
\end{equation}
In the special case of $\mathbf{z}_a=x+iy$ and $\mathbf{z}_b=x+i(y+y_j)$, namely $\mathbf{z}_a$ and $\mathbf{z}_b$ have the same $x$ and a shift $y_i$ in $y$, 
\begin{equation}
	\rho_{1/q}(\mathbf{z},y_j,\mathbf{z})=\sum_k \langle c_k^{\dagger}c_k \rangle_{1/q} |\phi^{c}_k(\mathbf{z})|^2 e^{ik y_j}.
\end{equation}
Since $\langle c_k^{\dagger}c_k \rangle_{1/q}$ is non-zero over a contiguous, finite and known range $k \in [-\frac{2\pi}{L_y} \frac{q(N-1)}{2},\frac{2\pi}{L_y} \frac{q(N-1)}{2}]$, the summation over $k$ can be restricted to this range without any uncertainty. 
Then the above relation could be explained as a discrete Fourier transformation from momentum space $k$ to real space conjugate $y$. The inverse transformation has the following form
\begin{equation}
\label{eq:a6}
	\langle c_k^{\dagger}c_k \rangle_{1/q}|\phi^c_k(\mathbf{z})|^2 =\frac{1}{N_{orb}}\sum_{j=0}^{q(N-1)}e^{-iky_j}\rho_{1/q}(\mathbf{z},y_j,\mathbf{z})
\end{equation}
where $y_j=\frac{L_y}{N_{orb}}j$ and $N_{orb} = qN-q+1$ is the number of orbits. Note that Eq.~(\ref{eq:a6}) is only true for $ -\frac{q(N-1)}{2} \leq m \leq  \frac{q(N-1)}{2}$. In principle Eq.~(\ref{eq:a6}) is valid for any value of $z$, but practically the resulting uncertainty in the occupation number will be a minimum when $r$ is near the maximum in $|\phi^{c}_k(\mathbf{z})|^2$ which occurs at $z\sim m\gamma l_B$. We evaluate the occupation number by integrating Eq.~(\ref{eq:a6}) over $\mathbf{z}$ to get,
\begin{equation} \label{occk}
	\langle c_k^{\dagger}c_k \rangle_{1/q}=\frac{1}{N_{orb}}\sum_{j=0}^{N_{orb}-1}e^{-iky_j}\rho_{1/q}(y_j)
\end{equation}
where $\rho_j= \rho_{1/q}(y_j)=\int d^2z \rho_{1/q} (\mathbf{z},y_j,\mathbf{z})$. Then the occupation at any $k$ (within the appropriate range) can be found after evaluating $\rho_j$ for all $j=0,\cdots, N_{orb}-1$. From Eq.~(\ref{eq:a3}), we have,
\begin{equation}
	\rho_j=\frac{N\int \prod_{i=1}^{N} d^2\mathbf{z_i}\psi^c_{1/q}(\mathbf{z_1-iy_j,\cdots z_{N}}) \psi_{1/q}^{c*}(\mathbf{z_1,\cdots z_{N}} ) }{\int \prod_{i=1}^{N} d^2\mathbf{z_i}|\psi^c_{1/q}|^2 }
\end{equation}
Ignoring the normalization factor, the Eq.~(\ref{eq:a1}) becomes,
\begin{equation}
	\psi^c_{1/q}(\mathbf{z_1-iy_j},\mathbf{z_2},\cdots)=\psi^c_{1/q}(\mathbf{z_i})Z_1(y_j,\mathbf{z})
\end{equation}
where
\begin{equation}
	Z_b(y_j,\mathbf{z})=\prod_{k\neq b}\frac{(e^{\gamma(z_b-iy_j)}-e^{\gamma z_k})^q}{(e^{\gamma z_b}-e^{\gamma z_k})^q}e^{i\gamma \frac{q(N-1)}{2}y_j}
\end{equation}
thus we have
\begin{equation}
	\rho_j=\frac{N\int \prod_{i=1}^{N} d^2\mathbf{z_i}|\psi^c_{1/q}|^2 Z_1(y_j,\mathbf{z})}{\int \prod_{i=1}^{N} d^2\mathbf{z_i}|\psi^c_{1/q}|^2 }
\end{equation}
In the final, the $\rho_j$ can be expressed as
\begin{equation}
	\rho_j=\frac{\int \prod_{i=1}^{N} d^2\mathbf{z_i}|\psi^c_{1/q}|^2 \sum_{b=1}^{N} Z_b(y_j,\mathbf{z})}{\int \prod_{i=1}^{N} d^2\mathbf{z_i}|\psi^c_{1/q}|^2 }
\end{equation}
where we have symmetrized $Z_b$ over all particles index to increase the rate of convergence without loss of generality. The above expression can be evaluated through Metropolis sampling with a high accuracy. Then we can obtain the average occupation number of $1/q$ Laughlin state on cylinder after going back to Eq.~(\ref{occk}). In the similar scheme, we can obtained the occupation numbers for other FQH states, such as Moore-Read Pfaffian state and two-component Halperin 331 state. The technical details for these states are in the Appendix~\ref{sec:app_MR} and  \ref{sec:app_331}. 

\begin{figure}[ht]        
\center{\includegraphics[width=8.5cm]  {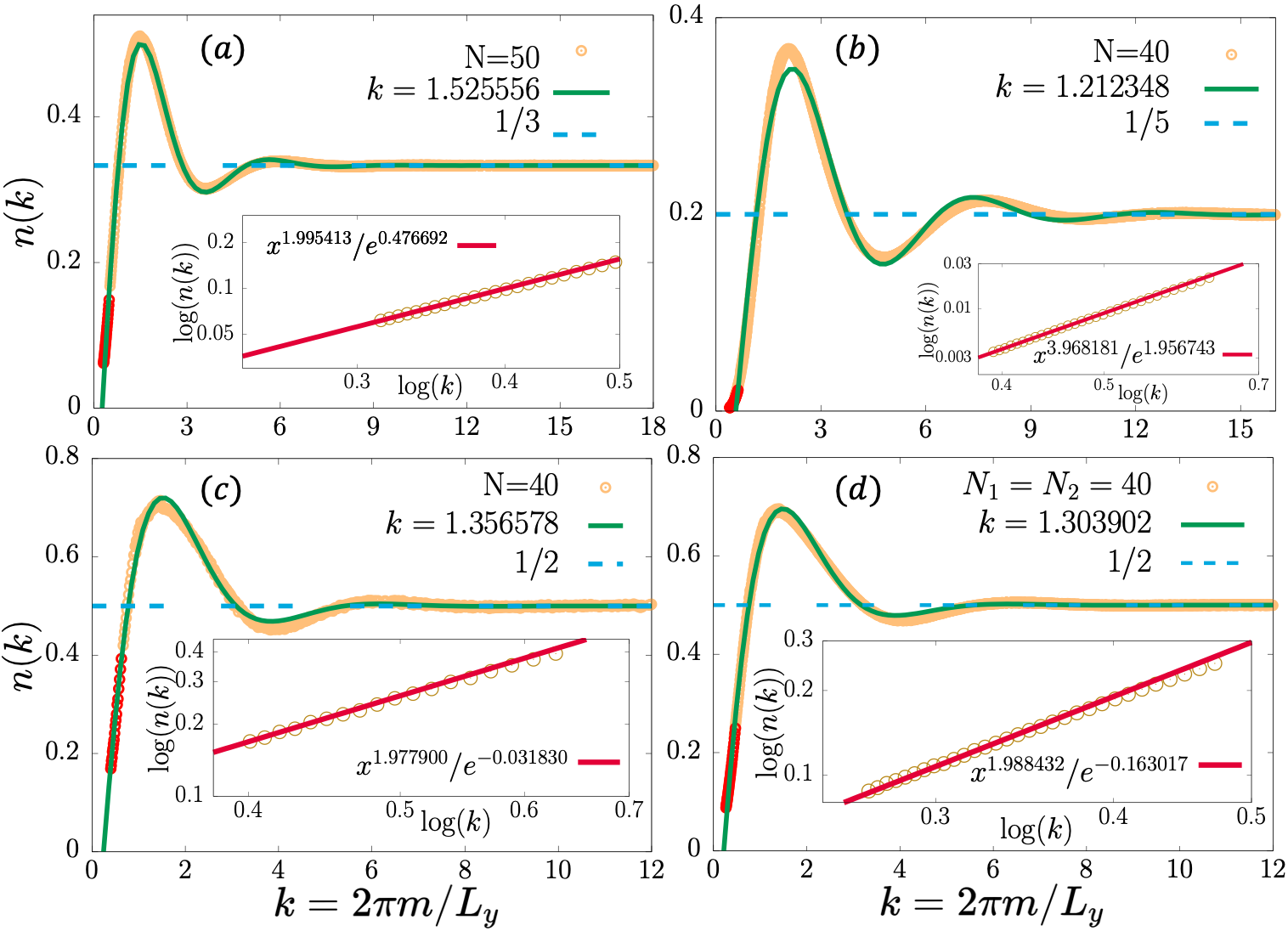}}        
\caption{\label{figoccall} Occupation numbers at the edge of (a) $\nu=1/3$ state for 50 particles, (b) $\nu=1/5$ state for 40 particles, (c) $\nu=5/2$ MR state for 40 particles and (d) $331$ state for 80 particles(40 in each layer).   The inset plots are the linear fit in logarithmic scale for the data near the Fermi point(labelled in red circles in $n(k)$). The slopes are $r_{1/3}=1.99541\pm 0.008402$, $r_{1/5}=3.96818\pm 0.03806$, $r_{MR}=1.9779\pm 0.03377$ and $r_{331}=1.98843\pm 0.04556$ which are exactly the same as predicted in CBT.   } 
\end{figure}

With the occupation numbers for large system, including $\nu = 1/3,1/5$ Laughlin states, $\nu=5/2$ Moore-Read and Halperin $331$ state, we verify the behavior of $n_k$ near the edge with comparing to the chiral boson theory in a high accuracy.  The magnetoroton minimum could also be fitted in a large range. Specifically, for $1/5$ and $331$ state, since the size of Hilbert space is extremely large in exact diagonalization. The bulk density is difficult to reach the uniform density $\rho = \nu/2\pi l_B^2$ and thus many of the physical quantities are obscured by finite size effects. Fig.~\ref{figoccall} shows half of the occupation numbers for these states due to the central symmetry.  The occupation numbers are plotted as a function of the wave vector $k=\frac{2\pi m}{L_y} $ rather than orbital index $m$. By properly choosing the Fermi points~\cite{Park} to assure there are $N/\nu$ orbitals between two Fermi points, i.e., the momentum of the first non-vanishing occupation number is $m_0=3/2$ for $\nu = 1/3, 5/2$ and $331$ states and $m_0=5/2$ for $\nu = 1/5$ state, the data for different circumferences $L_y$ (The $L_y$ takes value in the range $(a)[19l_B,30l_B]$, $(b)[25l_B,40l_B]$, $(c)[15l_B,23l_B]$ and $(d)[20l_B,35l_B]$ to make sure the two edges are well separated) collapses into a perfectly smooth curve which manifests the universality of the FQH edge. Since we now have much more data near the Fermi points and no breakpoints as comparing to the results from Jack polynomials.~\cite{Park} The $n(k)$ near the edge clearly demonstrates that the FQH edge is described by the CBT with $n_k \propto k^r$ in which $r = \nu^{-1}-1$ for Laughlin states. Thus we take the linear fit of $\log n(k)$ versus $\log k$ with the first non-vanishing occupation number of different $L_y$. For the four FQH states we considered, the CBT predicts their exponents being $r = 2, 4, 2, 2$ respectively. Our simulation gives these fitting values as $r=1.99541 \pm 0.008402,3.968181 \pm 0.03806,1.9779\pm 0.03377,1.98843 \pm 0.04556$ as shown in the insets of each figure in Fig.~\ref{figoccall}. They are exactly the same as the expected value within the statistical error.

On the other hand, as in Ref.~\onlinecite{Shibata},  we fit the oscillations of the occupation numbers by $f_\nu(x)=c_\nu \exp(-x/\epsilon_\nu)\cos(k_\nu x+\theta_\nu)+\nu$. It was claimed that $k_\nu$ is in good agreement with the wave number of the bulk magnetoroton minimum and $\epsilon_\nu$ is proportional to the bulk excitation gap. The density oscillation at the edge reflecting the bulk excitation is a good example of the bulk-edge correspondence in topological ordered phase. From our simulations of the model wave functions, the fitting parameters are $\epsilon_{1/3}=1.357$, $k_{1/3}=1.526$; $\epsilon_{1/5}=2.415$, $k_{1/5}=1.212$, $\epsilon_{5/2}=1.185$, $k_{5/2}=1.357$ and $\epsilon_{331}=1.08$, $k_{331}=1.304$ respectively.  Here we should note that our results are for the model wave functions which corresponding to the eigenstates of the model Hamiltonian, such as $V_1$ Haldane pseudopotential Hamiltonian for $\nu = 1/3$ Laughlin state. The realistic Coulomb interaction naturally gives different results, especially the energies. Comparing to the result of density matrix renormalization group with Coulomb interaction~\cite{Shibata}, the $k_{1/3}$ is quite close and $\epsilon_{1/3}$ is very different as expected since the wavefunction is quite close and energy should be different. Therefore, we expect that the magnetoroton minimums for other three FQH states($1/5$, MR and $331$) in Coulomb Hamiltonian are almost the same as the $k_\nu$ we obtained. 

\section{Topological Quantities from Momentum Polarization}
\label{sec3}
It is known that the quasihole mutual exchange in the FQH liquids contains rich information of its topological order. Suppose we have a quasihole on each edge of cylinder, the rotation along $y$ direction will not give any information since the rotational symmetry of this manifold. However, if one can rotate half of the cylinder (subsystem A) and keep another half (subsystem B) unchanged, the many-body wave function will have a phase containing the information of the quasihole in subsystem A. This phase is called momentum polarization,~\cite{HHTYZXLQ} which contains important topological quantities such as the Hall viscosity,~\cite{Read09} guiding-center spin, central charge and the topological spin (conformal dimension)  of the quasihole excitation.  Momentum polarization has previously been studied using the entanglement entropy in cylinder geometry~\cite{HHTYZXLQ} and modular transformation in torus geometry.~\cite{MPZRSKMFP, LDHWZ} It could also be studied in the entanglement spectrum at the bipartite boundary in the bulk and the intrinsic dipole moment from the density profile on the edge.~\cite{Park, LDHWZ}

Here we firstly employ the occupation numbers to calculate momentum polarization. It can be acquired by
\begin{equation}\label{MP}
     \langle \Delta M_A \rangle=\sum_{m\in A}m\langle n_m \rangle-M_A^0
\end{equation}
where $M_A^0$ just depends on the root occupation number, such as $1001001001\cdots$, $1000010000100001\cdots$, $1100110011\cdots$. Theoretically, the momentum polarization contains three leading terms as follows
\begin{equation}
	\langle \Delta M_A \rangle=\frac{\eta_H}{2\pi\hbar}L_y^2 -h_\alpha+\frac{\gamma}{24}
\end{equation}
where the first term is from the contribution of guiding-center Hall viscosity. The second term $h_{\alpha}=M_A^0-\bar{M_A}$ is called topological spin~\cite{MPZRSKMFP,HHTYZXLQ} or conformal spin of the elementary excitations which corresponds to quasihole sector $\alpha$ and depends on the position of the bipartition in the occupation space. It can be calculated for different model FQH states by using the root configuration pattern in the Jack polynomial description or the conformal field correlator of the quasihole operators as shown in the Appendix~\ref{sec:app_C}. The third leading term $\gamma=\widetilde{c}-\nu$ is the difference between the (signed) conformal anomaly ($\widetilde{c}=c-\bar{c}$) and the chiral charge anomaly (filling factor) $\nu$, which are the two fundamental quantum anomalies of the FQH fluids. The theoretical values are as follows: $c=1$ for Laughlin states, $c=3/2$ for the $5/2$ Moore-Read state and $c=2$ for the bilayer $331$ state, and all chiral states have $\bar{c}=0$. Notice that $\gamma$ vanishes in integer quantum Hall states, which are topological trivial. 

In the case of FQH fluid, the edge density deviates from the uniform density $\nu/2\pi l_B^2$ due to the electron-electron correlation. This nonuniform occupation distribution gives a quantized dipole moment $p_x$, which is related to the guiding-center Hall viscosity (the expectation value of area-preserving deformation generators).~\cite{Park}$^{,}$\cite{LDHWZ} The essential physics here is the intrinsic dipole momentum coupling with the gradient of the electric field from the Coulomb interaction and the confining potential. This coupling results in an electric force which is balanced by the guiding center Hall viscosity $\eta_H$. Moreover, the guiding center Hall viscosity was found to have a relation to a topological quantity named as the guiding-center spin. $\eta_H^{ab} = -\frac{\hbar}{4\pi l_B^2} \frac{s}{q} g^{ab}$ where $g^{ab}$ is the guiding-center metric in Haldane's geometric description of FQH liquid~\cite{HaldanePRL11} and the guiding-center spin $s$ coupling with the curvature gives the topological shift on sphere. Finally, we have the relation 
\begin{equation}
\label{hallvis}
	 \eta_H=-\frac{p_x}{L_y}B=\frac{\hbar}{4\pi l_B^2}\frac{-s}{q}
\end{equation}
where $B$ is the strength of magnetic field and $q$ is the flux quantua number attached by a ``composite boson" which is made of $p$ particles with $q$ flux quanta for $\nu = p/q$. After a simple substitution, we have
\begin{equation}
\label{eq:7}
	\langle \Delta M_A \rangle=-\frac{1}{2}(\frac{L_y}{2\pi l_B})^2\frac{s}{q}-h_\alpha+\frac{c-\nu}{24}
\end{equation}

The edge dipole moment per length could be calculated from the occupation numbers as follows,
\begin{equation}
	\frac{p_x(k)}{L_y}=-\frac{e}{2\pi}\int_0^k dk' k'l_B^2 [n(k')-\nu].
\end{equation}
and for finite $L_y$, comparing to Eq.~(\ref{eq:7}), the integration is approximated by the sum with corrections 
\begin{equation}
\label{px}
	\frac{p_x(k)}{L_y}=-\frac{2\pi l_B^2 e}{L_y^2}(\sum_{m'}[n(m')-\nu]m'+ h_\alpha-\frac{c-\nu}{24}). 
\end{equation}
Here we should note that the origin paper of Eq. (37) in Ref.~\onlinecite{Park} does not have the correction terms and the difference is also discussed detailedly in a recent work.~\cite{LDHWZ} It is due to the  equivalence of intrinsic dipole moment and momentum polarization, which can be considered as the same topological quantity. On the other hand, we point out that this may also be the reason why Fig. 16-20 in~\onlinecite{Park} is less convergent. A slight shift between theoretical and numerical values is clearly observed there which is clearly not a finite size effect.  The quantities $s, h_{\alpha}, c$ contains very rich information. Guiding center spin $s$ is related to the non-dissipative response of the metric perturbation.~\cite{LHZLDNSFDMHWZ} Topological spin $h_{\alpha}$ and central charge $c$ are the elements of modular-$T$ matrix, which is the unitary transformation of the ground state manifold under modular transformation.~\cite{EKVXGW}

In our simulation process, we use a self-consistent test method to determine these topological quantities, namely we set the other quantities at their respective theoretical values when calculating one of them. For example, while calculating the central charge $c$, the guiding center spin and topological spin are predetermined by their CFT values and thus 
\begin{equation}
	c=24\langle \Delta M_A \rangle+24\cdot \frac{L_y^2}{8\pi^2}(-\frac{1}{3})+24h_\alpha +\frac{1}{3}
\end{equation}
for $\nu = 1/3$ Laughlin state. The results of the topological spin strongly depend on how many quasiparticle the subsystem has. This could be tuned by shifting the bipartite position in the root configuration of the Jack description.~\cite{Bernevig08a,Bernevig08b} Basically, there are three topological sectors for $1/3$ Laughlin state with root $010010\cdots010010$. One is the equal bipartition with $\cdots10|01\cdots$ named vacuum cut in which case the subsystem of $N_A$ particles exactly occupy $N_A/\nu = 3N_A$ orbitals and thus no quasiparicle (quasihole) excitation. If one more (less) orbital is allocated to subsystem, such as  $\cdots100|1\cdots$ ( $\cdots1|001\cdots$), a quasihole (quasiparticle) is created in the left subsystem. The different bipartitions and their corresponding topological spins in other FQH states are discussed in  Appendix~\ref{sec:app_C}. Finally, for a specific system with fixed number of electrons, we calculated these topological quantities by varying the aspect ratio of cylinder, or changing the $L_y$ with keeping the area invariant. Therefore, each $L_y$ gives one set of results as shown in all of the following results. 
\begin{figure}[ht]        
\center{\includegraphics[width=8.5cm]  {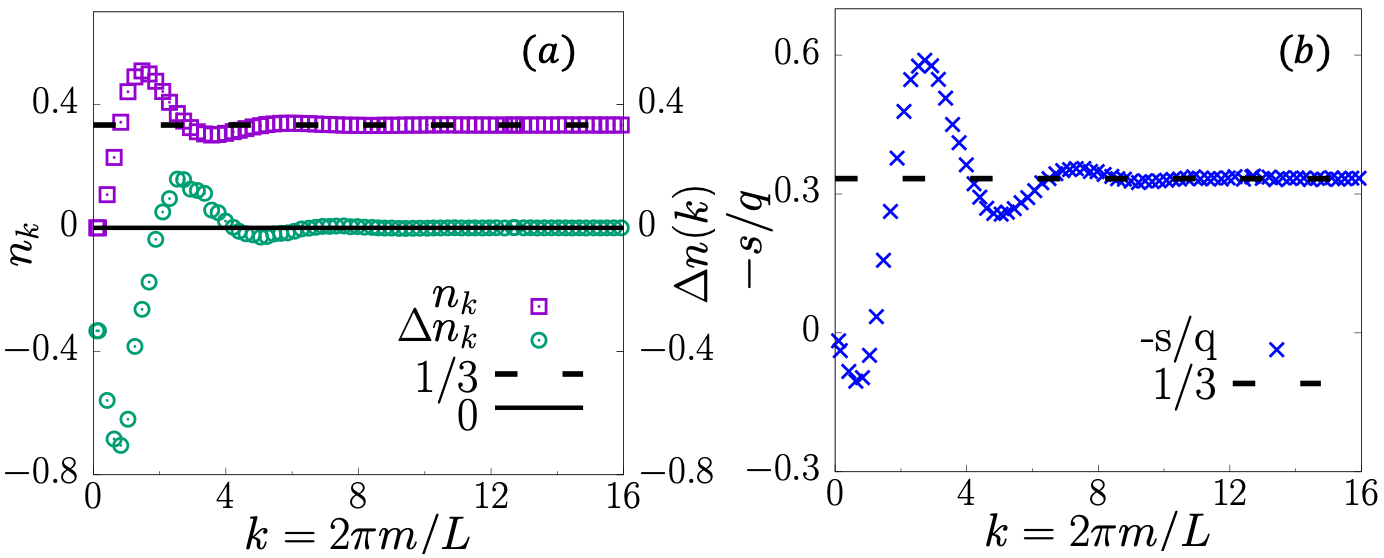}}      
\caption{\label{figgslau3}  (a) $\Delta n(k)$ and $-s/q$  in half of the cylinder for $1/3$ Laughlin state with 50 particles. The $\Delta n(k)$ converges to $0$ verifies the electric neutrality condition and convergence of the simulation. (b) The $-s/q$ converges perfectly to the expected value and thus $s = -1$.}
\end{figure}

\begin{figure}[ht]        
\center{\includegraphics[width=8.5cm]  {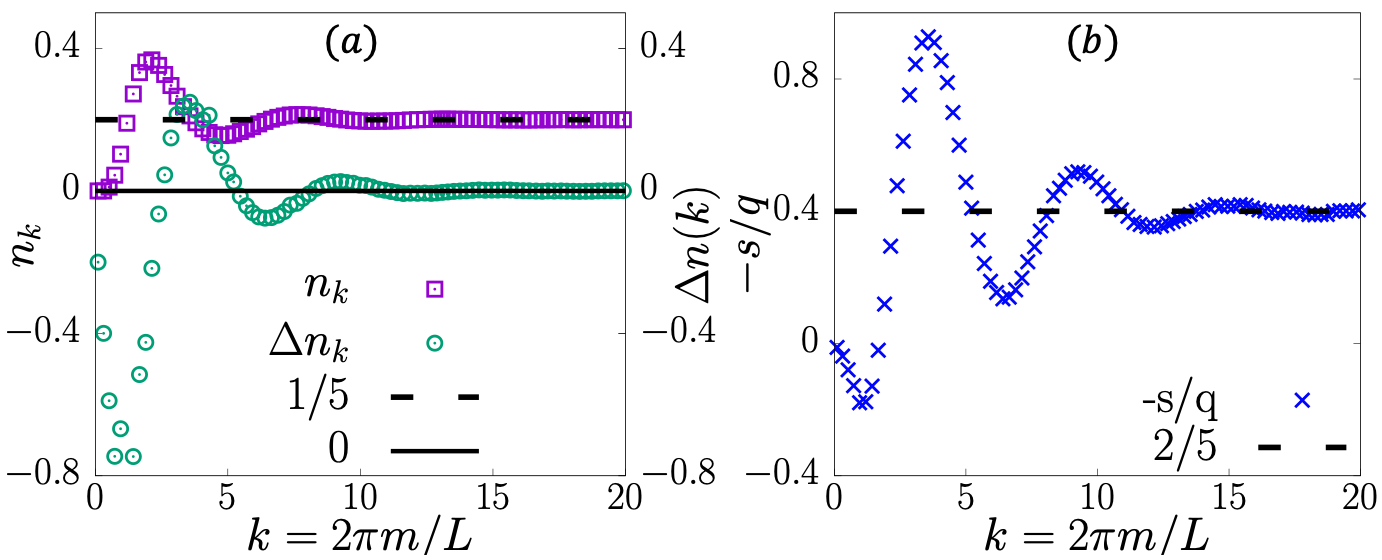}}      
\caption{\label{figgslau5} Same as Fig.~\ref{figgslau3} for $1/5$ with 40 particles. The guiding center spin converges to $s = -2$.}
\end{figure}

\begin{figure}[ht]        
\center{\includegraphics[width=8.5cm]  {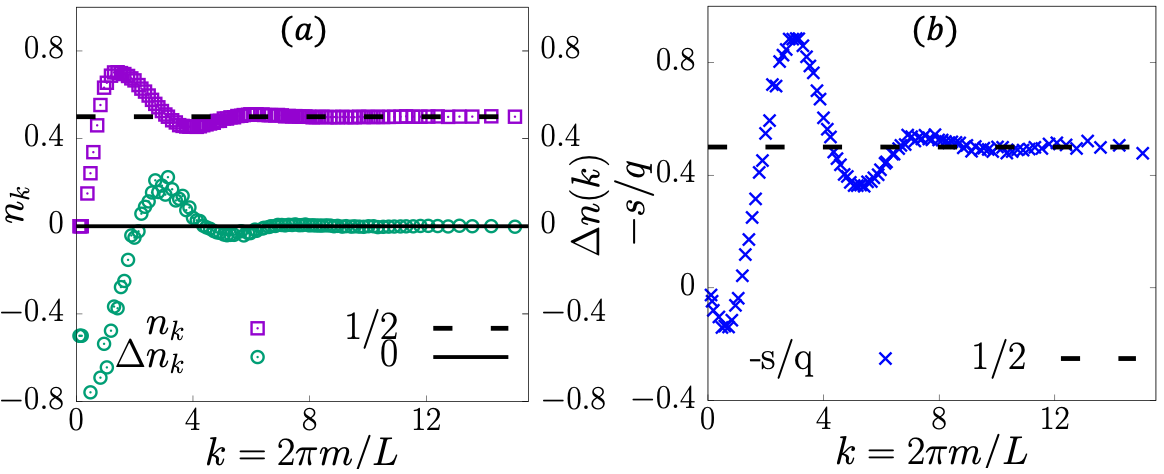}}        
\caption{\label{figgsmr} Same as Fig.~\ref{figgslau3} for MR state with 40 particles. The guiding center spin converges to $s = -2$. } 
\end{figure}

\begin{figure}[ht]        
\center{\includegraphics[width=8.5cm] {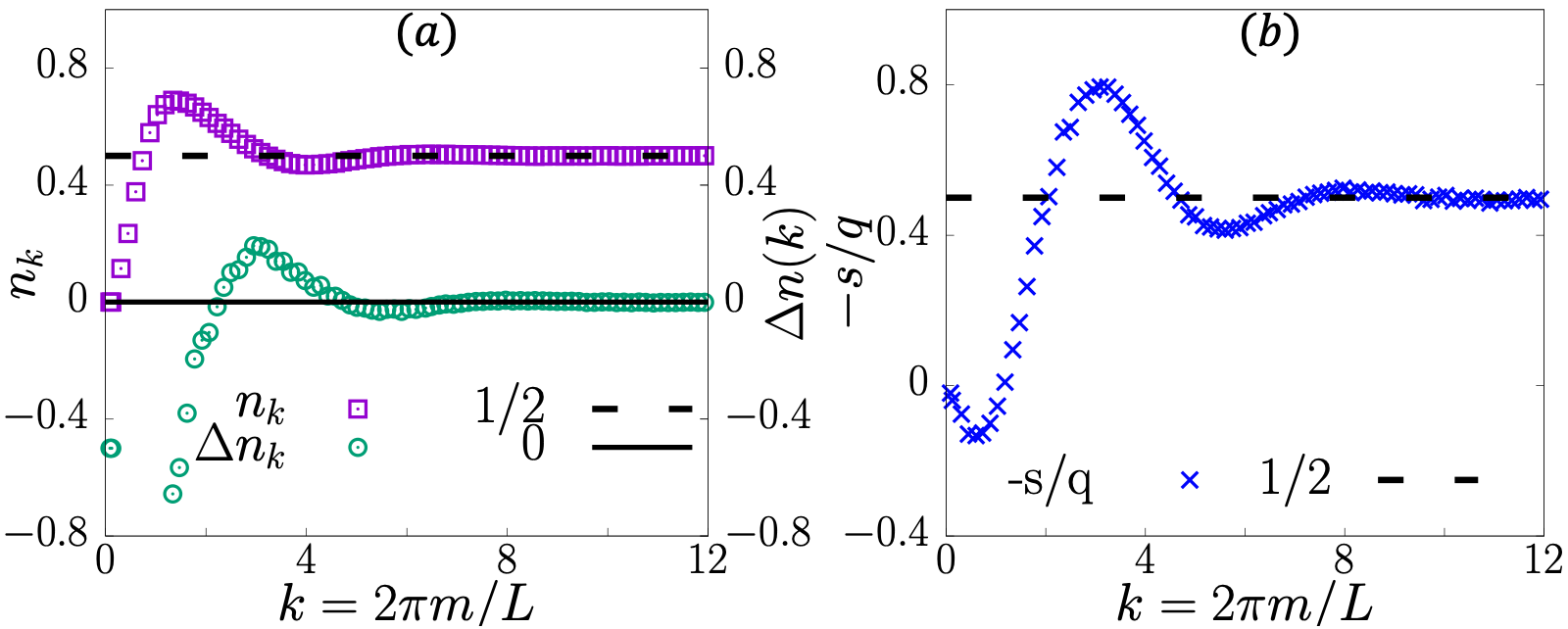}}        
\caption{\label{figgs331} Same as Fig.~\ref{figgslau3} for 331 state with 80 particles. The guiding center spin converges to $s = -2$.  } 
\end{figure}

 Combining Eq.~(\ref{hallvis}) and (\ref{px}), considering the contribution of central charge $c$ and topological spin $h_\alpha$~\cite{HHTYZXLQ} to the guiding-center spin, and discretizing the momentum, we have
\begin{equation}
	-\frac{s}{q}=\frac{8\pi^2}{L_y^2}(\sum_{m_1}(n(m_1)-\nu)m_1+ h_\alpha-\frac{c-\nu}{24} )
\end{equation}
Before calculating guiding-center spin, we need to validate the Luttinger's sum rule,~\cite{J.M. Luttinger} i.e. charge neutral conditions 	$\sum_{m_1}[n(m_1)-\nu]=0$.
Our numerical results are shown in Fig.~\ref{figgslau3}-\ref{figgs331}. Firstly we verify the Luttinger's sum rule, the difference $\Delta n_k$ between the occupation number and the uniform occupation $\nu$ converges to zero. Then, the $-s/q$ converges to $1/3, 2/5, 1/2, 1/2$ respectively, which gives us the guiding-center spin for $1/3, 1/5$ Laughlin state, $5/2$ MR state and 331 state as $s = -1,-2,-2,-2$ respectively. 

\begin{figure}      
\center{\includegraphics[width=8.5cm]  {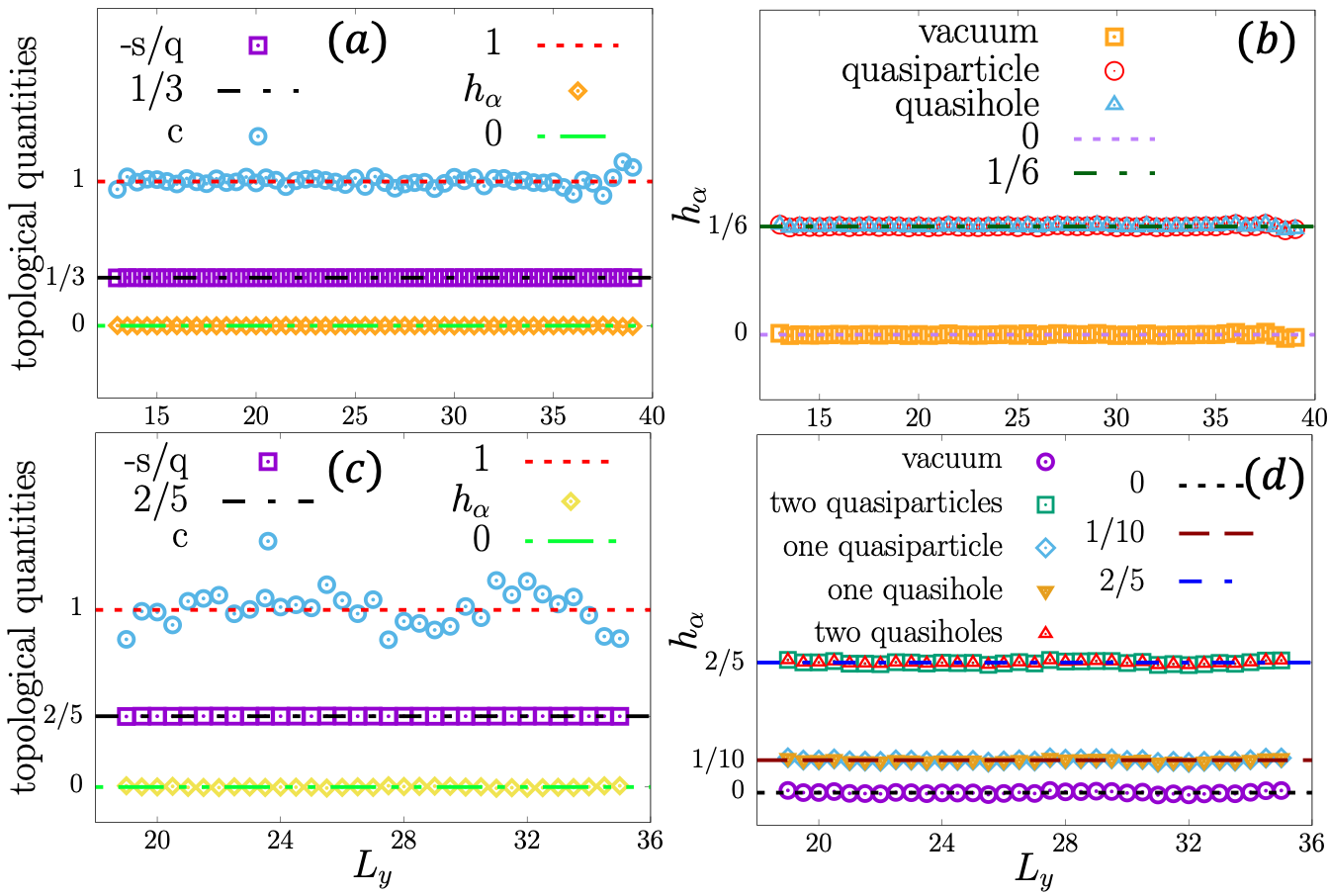}}        \caption{\label{figmplau} Topological quantities for $\nu = 1/3$ and $\nu = 1/5$ Laughlin states.
In (a) and (c), we present the result of $-s/q$, $c$ and $h_\alpha$ in the vacuum cut. (b) and (d) show the $h_\alpha$ for different bipatitions. The corresponding theoretical results are labelled by horizontal  dash lines.} 
\end{figure}

\begin{figure}       
\center{\includegraphics[width=8.5cm]  {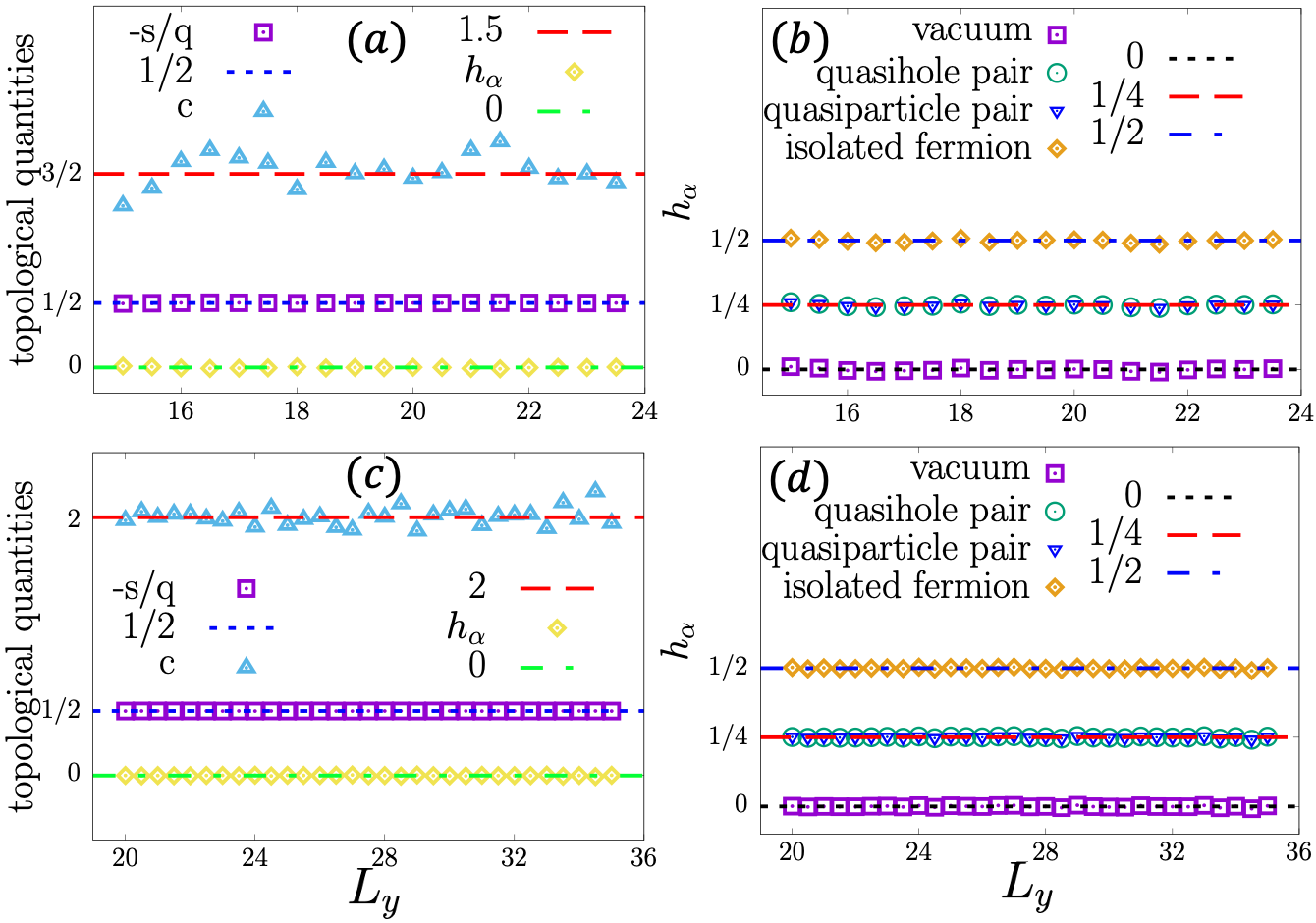}}        \caption{\label{figmpmr} Same as Fig.~\ref{figmplau}. Topological quantities for MR and 331 states. Here the quasihole/quasiparticle is $e/2$ charged since one orbit is shifted in the occupation number configuration.}
\end{figure}

\begin{figure}       
\center{\includegraphics[width=8.5cm]  {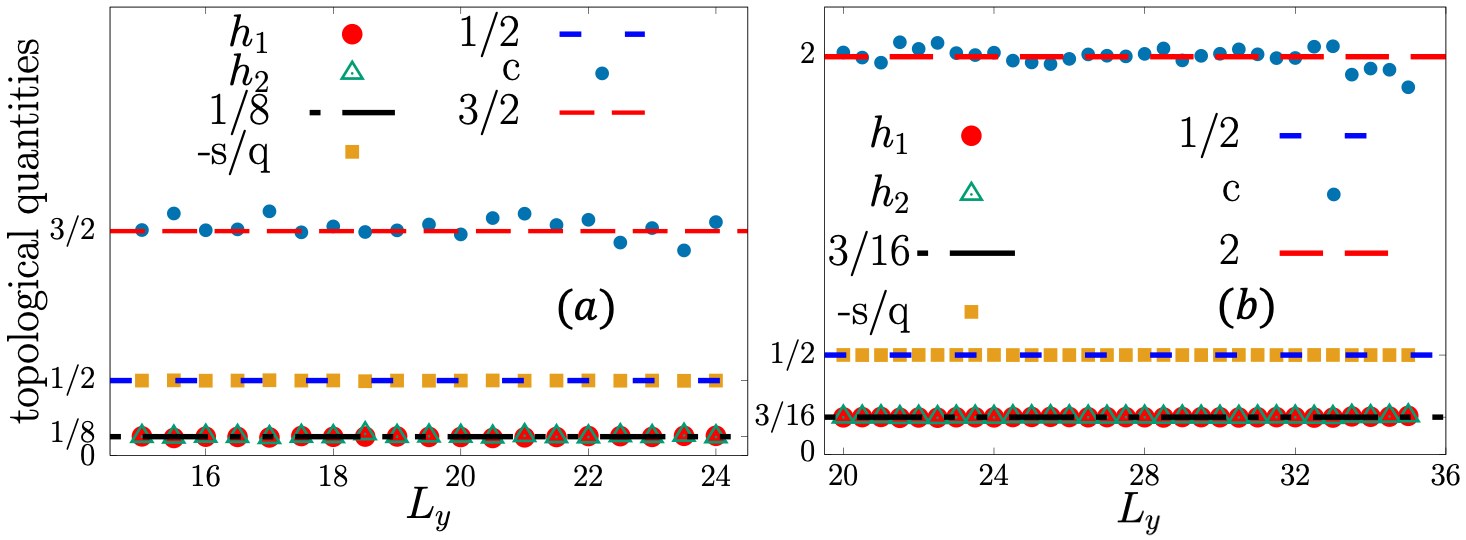}}        \caption{\label{figmpmrqh} Topological quantities for (a) Moore-Read state and (b) 331 state with $e/4$ quasihole excitation on the edge. }
\end{figure}

Now we go back to Eq.~(\ref{eq:7}) to calculate the other topological quantities. We extract these topological quantities numerically of different FQH states, the results are shown in Figure~\ref{figmplau}-\ref{figmpmrqh}. 
First of all, we observe that the Monte Carlo simulation of the large systems indeed gives us much more accurate topological quantities of the FQH states. Those values are in good agreement with the theoretical predictions from the CFT. For example, we get $c=1,1,3/2,2$, $s=-1,-2,-2,-2$ for $1/3,1/5$ Laughlin states, Moore-Read state and $331$ state respectively.  As for topological spin, all theoretical predictions are presented in Appendix~\ref{sec:app_C}. In comparison with the previous study by matrix product state (MPS) with a low truncation level,~\cite{Park} the accuracy of our method is prominent and the computational cost is effective. Especially for the $1/5$ Laughlin state, it is found that the convergence of these topological quantities are very slow comparing to the $1/3$ Laughlin state. For example, Fig.~\ref{figmplau}(c) shows that the central charge has apparent larger fluctuations than that in Fig.~\ref{figmplau}(a). The slow convergence in $1/5$ is clearly shown in the scope of the edge density fluctuation in Fig.~\ref{figgslau5}.

From Fig.~\ref{figgsmr}, Fig.~\ref{figgs331} and  Fig.~\ref{figmpmr}, we can see that except the central charge $c$, all the other topological quantities are the same for Moore-Read state and $331$ state. It is known that their $e/4$ quasihole excitations are very different in the anynonic statistics. The $e/4$ quasihole in the Moore-Read state is non-Abelian since it contains Majorana mode and that in the $331$ state is trivial Abelian. Here we model these quasiholes at the edge of the cylinder in Monte Carlo (details are shown in Appendix~\ref{sec:app_qh}) and calculate their topological quantities as shown in Fig.~\ref{figmpmrqh}. From which we find that the central charge and guiding-center spin are the same as the ground state. However, for topological spin of the $e/4$ quasihole, the numerical results show that $h_\alpha = 1/8$ for Moore-Read state and $h_\alpha = 3/16$ for 331 state.
These values are in good agreement with their theoretical predictions as shown in Appendix~\ref{sec:app_C} which demonstrates their different topological properties. 

\section{Edge Green's function}
\label{sec4}
Owing to the existence of gapless edge states in the FQH liquids with open boundaries, current exists between two contacts connected by an edge channel, as electrons can be injected into or removed from the FQH edge with costing zero energy. The standard theory for the FQH edge physics is the $\chi LL$ theory.~\cite{XGW,XGWQN} The theory predicts that a FQH droplet exhibits a power-law behavior in the electric current-voltage characteristics ($I \propto V^{\alpha}$) when electrons are tunneling through a barrier into the FQH edge from a Fermi liquid.~\cite{XGW,XGWQN,AMCLNPKWW,AMC} Generally speaking, $\alpha$ is also a topological quantity which is related to the topological order of the FQH liquid and immune from the perturbations. For the celebrated $\nu = 1/3$ Laughlin state, the $\chi LL$ theory predicts a tunneling exponent $\alpha = 3$ though it has controversial in realistic system as we mentioned in the introduction.  The $\alpha$ measured in experiments is sample dependent with a value mostly smaller than 3.~\cite{AMCLNPKWW,MGDCTLNPKWWAMC,AMCMKWCCCLNPKWW,AMC}  One of the possible causes of this discrepancy is existence of counterpropagating edge modes, which result from edge reconstruction.~\cite{Chamon94,XWFEEHR,VJGEVT,XWKYEHR,KY}  The $\chi LL$ theory~\cite{ WXG1,WXG2} also predicts the $\alpha = 5$ for $1/5$ Lauglin state, $\alpha = 3$ for both of the Moore-Read and $331$ states. The related experimental and theoretical values for Laughlin states are in Ref.~\onlinecite{WXG3,MGDCTLNPKWWAMC,XLCDMAKLNPKWW,HLFPJWPJSLXLNPKWMAKXL}. 

Numerically, we can obtain $\alpha$ by calculating the electron edge Green's function which is the electron propagator along the edge of the FQH droplet. The scaling behavior of edge Green's function has been studied in disk geometry.~\cite{SSMJKJ1,XWFEEHR,XWKYEHR} As we claimed previously, the disk geometry has inhomogeneous Landau orbital space and thus the edge density profile is always incomplete in small system size. Another problem is the edge distance is limited by the circumference of disk and the scaling behavior suffers strong finite size effects.  The Monte Carlo simulation in cylinder geometry could overcome these weaknesses since the lengthscale of the edge could be tuned by the aspect ratio. 
 In cylinder geometry, the edge Green's function can be defined as
\begin{equation} \label{edgegreencylinder}
	G_{edge}(|\vec{z}-\vec{z'}|)=\frac{N\int \prod_{j=1}^{N-1} d^2 z_j \psi^{*} (\vec{z}, \{ \vec{z_j} \} ) \psi (\vec{z'}, \{ \vec{z_j} \} )}{\int \prod_{k=1}^N d^2 z_k \psi^{*}(\{ \vec{z}_k \}) \psi (\{ \vec{z}_k \}) }
\end{equation}
where $\vec{z}$ and $\vec{z'}$ are on the same edge of cylinder, and they have the same value of $x$ coordinates. In the limit of large distance ($|\vec{z}-\vec{z^{'}}| \gg 1$), the Green's function behaves as
\begin{equation}
	G_{edge}(|\vec{z}-\vec{z'}|) \sim |\vec{z}-\vec{z'}|^{-\alpha}
\end{equation}

From Appendix~\ref{sec:app_B}, the equal-time edge Green's function  on cylinder can be written as
\begin{equation}
	G_{edge}(|\frac{2}{\gamma} \sin(\frac{Y\gamma}{2})|)=\sum_k \frac{1}{\pi^{1/2}L_y} e^{-ikY} e^{-(X-k)^2} n_k
\end{equation}
The chord distance is $|\frac{2}{\gamma} \sin(\frac{Y\gamma}{2})|$ where $Y = |y_1 - y_2|$ is the arc length between $\vec{z}$ and $\vec{z'}$ on the surface of cylinder and $\gamma=2\pi/L_y=1/R$ is the inverse of the radius or the space between two continuous Landau orbits. 

\begin{figure}       
\center{\includegraphics[width=8cm]  {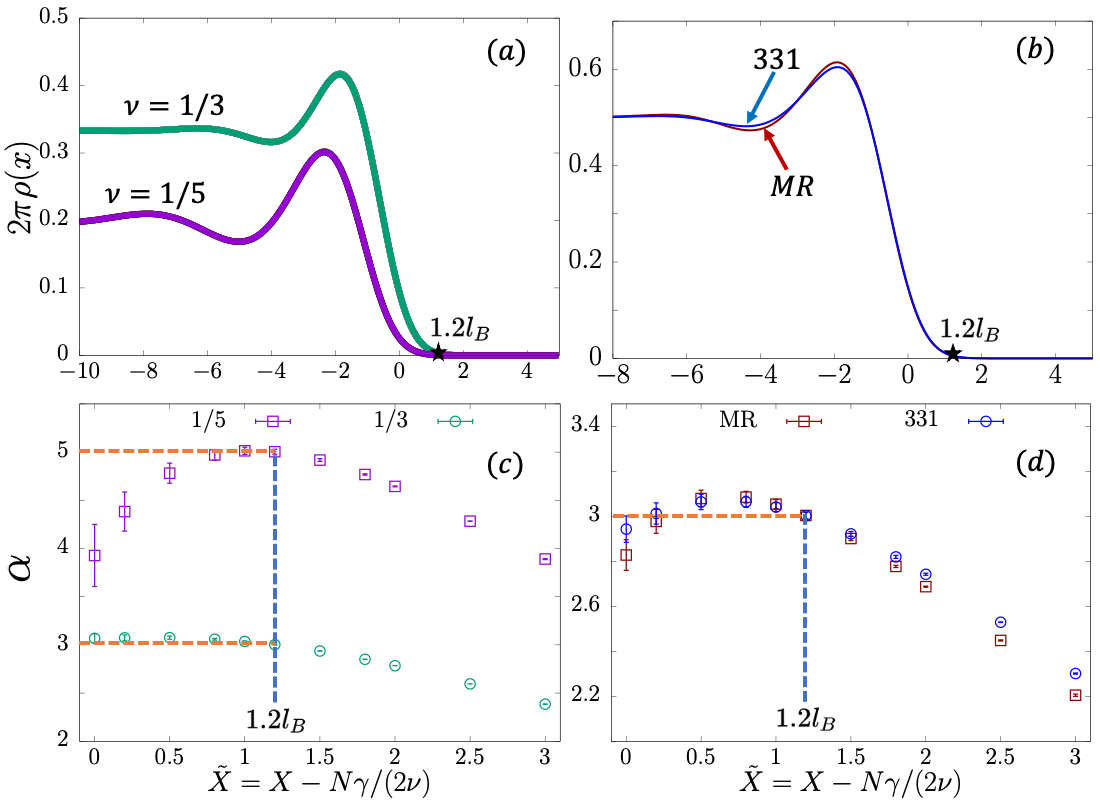}}  
\caption{\label{greenden} The density profiles (a)(b) and scaling exponents of the edge Green's functions at different $\tilde{X}$ (c)(d). Here the $x$-axis is set to $\tilde{X}=X-N\gamma/(2\nu)$, where $N\gamma/(2\nu)$ is the physical edge on the right. The theoretical predictions of the respective FQH states are marked as horizontal dashed lines.} 
\end{figure}

For a $N$-particle FQH liquid at filling $\nu$, the number of Landau orbits is $N/\nu$ and thus the length of the cylinder is $\frac{2\pi N}{\nu L_y} = N \gamma /\nu$. Two edges locate at $\pm N \gamma /2\nu$. As shown in Fig.~\ref{greenden}(a) and (b), we plot half of the density profile for the four states. Here we set the $x$ coordinate as $\tilde{X} =  X - N \gamma /2\nu$ and then the edge on the right side locates at the position of $\tilde{X}=0$. First of all, as we mentioned previously, the $1/5$ state has a deeper density oscillation than that of the $1/3$ state. Comparing the Moore-Read state and the 331 state with the same  electron number and $L_y$, the bulk density  are the same since both of them are candidates of the $\nu = 5/2$ FQH state. However, it is shown that the edge density has certain differences which demonstrates they belong to different topological phases, or have different topological quantities such as the central charge as that in previous section. For the Green's function along the edge, we fix the $x$ position and calculate Eq.~(\ref{edgegreencylinder}) in $y$ direction. Because the density profile always has a tail near the edge, we sweep the position of $\tilde{X}$ around $\tilde{X} = 0$. For each $\tilde{X}$, we calculate the Green's function and extrapolate the exponent $\alpha$ by the data of the large distance. The results are shown in Fig.~\ref{greenden}(c) and (d). Overall, we find the exponent $\alpha$ has a dependence on $\tilde{X}$. The interesting thing is that the values of $\alpha$ for all the four states reach to their respective theoretical value at around $\tilde{X} \simeq 1.2 l_B$ which indicated by a star in Fig.~\ref{greenden}(a) and (b). When $\tilde{X} > 1.2 l_B$, we find the $\alpha$ always decays and becomes smaller than the theoretical value. We understand this result as that the edge of the FQH liquid always has a width in order of one magnetic length $l_B$. The Luttinger liquid exponent has its exact value at the tail of the realistic edge where the electron density is close to zero as shown in Fig.~\ref{greenden}(a) and (b). This is acceptable since only the electrons at the tail of the edge are on the Fermi points and have gapless excitation. The electrons that are away from the Fermi points need a finite energy to excite and thus can not be strictly described as the gapless edge excitation, or $\chi LL$ theory.   
\begin{figure}       
\center{\includegraphics[width=8cm]  {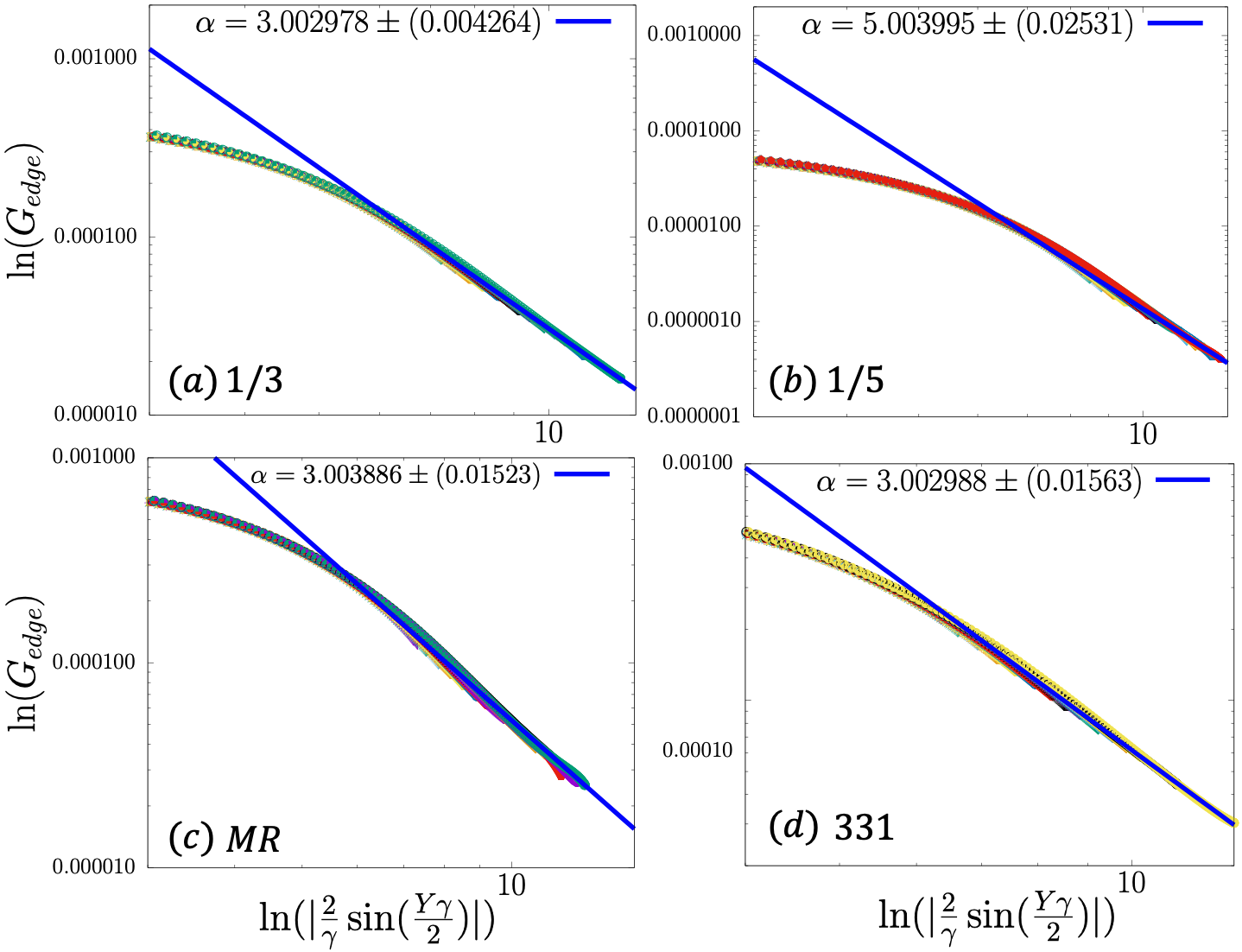}} 
\caption{\label{green} The edge Green's function at $\tilde{X}= 1.2l_B$ for (a) 1/3 Laughlin state, (b) 1/5 Laughlin state, (c) MR state and (d) 331 state. } 
\end{figure}

At the place of $\tilde{X} = 1.2 l_B$, we show the Green's function as a function of the chord distance in logarithmic plot in Fig.~\ref{green}. Similar to the density profile, the edge Green's function for different $L_y$ (aspect ratio) collapses into one curve with a small finite size fluctuation. The $\alpha$ displayed in the figure is obtained by fitting the data for large chord distance. The Moore-Read state and $331$ state share the same $\alpha$ which illustrates the electron tunneling, such as in the strong tunneling limit of quantum point contact (QPC) experiment, can not distinguish the two inequable states. However, since their $e/4$ quasiholes have different topological quantities as calculated in the momentum polarization, we expect the quasihole tunneling, such as in the weak tunneling limit of QPC could give their distinctions.

\section{Summaries and Discussions}
\label{sec:discussion}
 In this work, we have applied a Metroplis Monte Carlo method to calculate the electron occupation numbers of the Landau orbits for FQH model wave functions in cylinder geometry. We consider large systems with more than 40 electrons of the $\nu = 1/3, 1/5$ Laughlin states and two candidates for the $\nu = 5/2$ FQH states, namely the Moore-Read Pfaffian state and the Halperin bilayer 331 state. With smooth date near the edge, the full density profiles of the edge states are obtained and the chiral boson theory of the FQH edge has been verified in a high accuracy. 
 As a first inspection of the effectiveness of this method, we numerically determine the topological quantities via the dipole moment and momentum polarization calculations. The guiding-center spin, central charge and topological spin of the quasihole all exactly converge to their respective theoretical values. Notably, since the non-Abelian nature of the MR $e/4$ quasihole excitation, its topological spin is very different from its Abelian counterpart in 331 state. We model the $e/4$ quasihole excitation both these two states and identify their topological spins, which are consistent to the CFT predictions. With the occupation numbers of large system, another quantity we recalculated is the non-Fermi liquid behaviors of the electron Green's function along the edge. As sweeping the locations, we find only the electrons near the physical boundary have the  theoretically predicted $\alpha$. Therefore, we conclude that the $\chi LL$ theory is an idealized description of the boundary of the FQH liquid. This could be another possible mechanism that the $\alpha$ is not quantized and sample dependent in realistic experiments even in the absence of the edge reconstruction. This method could be easily generalized to other FQH states or the interface between different FQH states.
 
\acknowledgments
Z-X. Hu thanks H-H. Tu for helping us to explain the 331 state in CFT description. Y. Yang thanks C-X. Jiang for numerical skill discussions.
The Pfaffian polynomial calculation was implemented by using the algorithm in Ref.~\onlinecite{Michael Wimmer}. This work was supported by National Natural Science Foundation of China Grant No. 11974064 and 12147102, the Chongqing Research Program of Basic Research and Frontier Technology Grant No. cstc2021jcyjmsxmX0081, Chongqing Talents: Exceptional Young Talents Project No. cstc2021ycjh-bgzxm0147, and the Fundamental Research Funds for the Central Universities Grant No. 2020CDJQY-Z003.
 
\appendix
\begin{widetext} 

\section{Moore-Read State}
\label{sec:app_MR}
The Moore-Read Pfaffian model wave function in cylinder geometry could be written as
\begin{equation}
	\begin{split}
     |\psi^c_{5/2} \rangle &=\text{Pf}[M(e^{\gamma Z})] \prod_{j<k} [\exp(\gamma \mathbf{z}_j)-\exp(\gamma \mathbf{z}_k)]^2  e^{-\frac{1}{2}\sum_{i=1}^N x_i^2} e^{-\sum_i \frac{(2N-3)\gamma }{2}\mathbf{z_i}}
    \end{split}
\end{equation}
where $\text{Pf}[M(e^{\gamma Z})]$ is the Pfaffian polynomial of the antisymmetric matrix $M(e^{\gamma Z})$. For 4 particles, it is
\begin{equation}
	\begin{split}
		\text{Pf}[M(e^{\gamma Z})] &= \frac{1}{\exp(\gamma \mathbf{z_1})-\exp(\gamma \mathbf{z_2})} \frac{1}{\exp(\gamma \mathbf{z_3})-\exp(\gamma \mathbf{z_4})}-
		 \frac{1}{\exp(\gamma \mathbf{z_1})-\exp(\gamma \mathbf{z_3})}\frac{1}{\exp(\gamma \mathbf{z_2})-\exp(\gamma \mathbf{z_4})} \\
 		& +\frac{1}{\exp(\gamma \mathbf{z_1})-\exp(\gamma \mathbf{z_4})} \frac{1}{\exp(\gamma \mathbf{z_2})-\exp(\gamma \mathbf{z_3})}
	\end{split}
\end{equation}
\par\ In the Metroplis algorithm, we just need $|\text{Pf}(M(e^{\gamma Z}))|^2=\det (M(e^{\gamma Z}))$. Using the same calculation method as $1/q$ Laughlin state, we have 
\begin{equation}
	\rho_j=\frac{\int \prod_{i=1}^{N} d^2 z_i|\psi^c_{5/2}|^2 \sum_{b=1}^{N} Z_b(y_j,\mathbf{z})}{\int \prod_{i=1}^{N} d^2 z_i|\psi^c_{5/2}|^2 }
\end{equation}
where
\begin{equation}
\label{eq:a18}
		Z_b(y_j,\mathbf{z}) = \prod_{k\neq b}\frac{(e^{\gamma(\mathbf{z_b}-iy_j)}-e^{\gamma \mathbf{z_k}})^2}{(e^{\gamma \mathbf{z_b}}-e^{\gamma \mathbf{z_k}})^2} e^{i\gamma (N-3/2) y_j} \frac{\text{Pf}[M(e^{\gamma R(\mathbf{z_b} \rightarrow \mathbf{z_b}-i y_j)})]}{\text{Pf}[M(e^{\gamma Z})]} 
\end{equation} 
 where $y_j=\frac{L_y}{N_{orb}}j$ with $j=0 \cdots 2N-3$ and $k$ is from $-\frac{2\pi}{L_y} \frac{2N-3}{2}$ to $\frac{2\pi}{L_y} \frac{2N-3}{2}$. Similarly, we can obtain the average occupation number of $5/2$ MR state on cylinder by Metropolis sampling. We use the algorithm of Ref.~\onlinecite{Michael Wimmer} to implement the Pfaffian polynomial.

\section{Halperin $331$ State} 
\label{sec:app_331}
For the bilayer Halperin 331 state on cylinder, we assume there are $N_1(N_2)$ electrons in upper(lower) layer. The unnormalized wave function is
\begin{eqnarray}
\label{331eq1} 
		|\psi_{331}^c \rangle &=&\prod_{i<j,i,j\in N_1} (\exp(\gamma \mathbf{z}_i)-\exp(\gamma \mathbf{z}_j))^3  \prod_{k<l,k,l\in N_2} (\exp(\gamma \mathbf{z}_k)-\exp(\gamma \mathbf{z}_l))^3  \prod_{i\in N_1,k\in N_2} (\exp(\gamma \mathbf{z}_i)-\exp(\gamma \mathbf{z}_k))  \nonumber\\ 
		&&  e^{-\frac{1}{2}\sum_{i=1}^N x_i^2} e^{-\sum_{i\in N_1} \frac{\gamma }{2}(4N_1-3)\mathbf{z_i}} e^{-\sum_{i\in N_2} \frac{\gamma }{2}(4N_2-3)\mathbf{z_i}} 
\end{eqnarray}

where $N=N_1+N_2$ is total number of electrons. The total momentum is $\frac{3N_1(N_1-1)}{2}+\frac{3N_2(N_2-1)}{2}+N_1N_2$. When $N_1 = N_2$, the total momentum is $M_{tot} = \frac{N(2N-3)}{2}$ which is the same as that of the Moore-Read state. Each layer has filling $1/4$. Similarly, we have 
 \begin{equation}
	\rho_j=\frac{\int \prod_{i=1}^{N} d^2 z_i|\psi^c_{331}|^2 \sum_{b=1}^{N} Z_b(y_j,\mathbf{z})}{\int \prod_{i=1}^{N} d^2 z_i|\psi^c_{331}|^2 }
\end{equation}
where
\begin{equation}
	Z_b(y_j,\mathbf{z})=\begin{cases}
      \prod_{k\neq b,k\in N_1 }\frac{(e^{\gamma(z_b-iy_j)}-e^{\gamma z_k})^3}{(e^{\gamma z_b}-e^{\gamma z_k})^3} \prod_{k\in N_2}\frac{e^{\gamma(z_b-iy_j)}-e^{\gamma z_k}}{e^{\gamma z_b}-e^{\gamma z_k}} e^{i\gamma \frac{4N_1-3}{2}y_j},~\text{if}~z_b \in N_1. \\
      \prod_{k\neq b,k\in N_2 }\frac{(e^{\gamma(z_b-iy_j)}-e^{\gamma z_k})^3}{(e^{\gamma z_b}-e^{\gamma z_k})^3}
      \prod_{k\in N_1}\frac{e^{\gamma z_k}-e^{\gamma(z_b-iy_j)}}{e^{\gamma z_k}-e^{\gamma z_b}} e^{i\gamma \frac{4N_2-3}{2}y_j},~\text{if}~z_b \in N_2.
    \end{cases}
\end{equation}
and $y_j=\frac{L_y}{N_{orb}}j$.

\section{The Edge Green's Function on Cylinder}
\label{sec:app_B}
Since the single particle wave function is $\phi_k(z)=\frac{1}{\sqrt{\pi^{1/2} L_y}} e^{iky}e^{-(x-k)^2/2} $, the edge Green's function can be transfer to 
\begin{equation}
	\langle \phi^{\dagger }(\vec{z_1}) \phi (\vec{z_2}) \rangle =\sum_k \frac{1}{\pi^{1/2}L_y} e^{ik(y_2-y_1)} e^{-(x_1-k)^2/2} e^{-(x_2-k)^2/2} \langle a_k^{\dagger} a_k \rangle
\end{equation}
the coordinates $\vec{z_1}$ and $\vec{z_2}$ are chosen with the same position $x_1 = x_2 = X$ near the edge and have a shift in $y$ direction $Y=y_1-y_2$. So the chord distance could be expressed as $|\frac{2}{\gamma} \sin(\frac{Y\gamma}{2})|$. In the final, the edge Green's function on the cylinder could be calculated by the occupation numbers as
\begin{equation}
	\langle \phi^{\dagger }(\vec{z_1}) \phi (\vec{z_2}) \rangle=\sum_k \frac{1}{\pi^{1/2}L_y} e^{-ikY} e^{-(X-k)^2} n_k.
\end{equation}

\section{Topological Spin}
\label{sec:app_C}
The topological spin $h_\alpha$ could be calculated from the root configuration in the occupation space as
\begin{equation}
	h_\alpha=M_A^0-\bar{M}_A.
\end{equation}
The subscript $\alpha$ represents different topological sectors and depends on the location of the bipartition for subsystem A. $\bar{M_A}=-\frac{1}{2}\nu m_F^2$ is the total momentum for the subsystem A with uniform occupation density $\nu$, where $m_F$ is the orbital number in A.

Taking $4$-particle system as an example, for $1/3$ Laughlin state, there are three topological sectors, vacuum cut sector: 010010$|$010010, quasiparticle cut sector: 01001$|$0010010 and quasihole cut sector: 0100100$|$10010. So we have
\begin{equation}
    \begin{split}
        &010010|010010~~~h_\alpha=0 \\
		&01001|0010010~~~h_\alpha=1/6 \\
		&0100100|10010~~~h_\alpha=1/6
    \end{split}
\end{equation}
where we only consider subsystem A on the left. The first momentum near the cut is $-1/2$. For example, for quasiparticle cut sector, $M_A^0=-\frac{1+7}{2}=-4$, and $\bar{M_A}=\frac{1}{3}(-\frac{1+3+5+7+9}{2})=-\frac{25}{6}$, so $h_\alpha=-4+\frac{25}{6}=\frac{1}{6}$.

For $1/5$ Laughlin state, there are five topological sectors: vacuum cut: 0010000100$|$0010000100, two quasiparticles cut: 00100001$|$000010000100, one quasiparticle cut: 001000010$|$00010000100, one quasihole cut: 00100001000$|$010000100, two quasiholes cut: 001000010000$|$10000100. So we have
\begin{equation}
	\begin{split}
		&0010000100|0010000100~~~h_\alpha=0 \\
		&00100001|000010000100~~~h_\alpha=2/5 \\ 
		&001000010|00010000100~~~h_\alpha=1/10 \\ 
		&00100001000|010000100~~~h_\alpha=1/10 \\
		&001000010000|10000100~~~h_\alpha=2/5
	\end{split}
\end{equation}
Then for MR state, there are four topological sectors: vacuum cut: 0110$|$0110, isolated fermion cut: 01$|$100110, $e/2$ quasiparticle  cut: 011$|$00110 and  $e/2$ quasihole  cut 01100$|$110. We have
\begin{equation}
	\begin{split}
		&0110|0110~~~h_\alpha=0 \\
		&01|100110~~~h_\alpha=1/2 \\ 
		&011|00110~~~h_\alpha=1/4 \\ 
		&01100|110~~~h_\alpha=1/4 
	\end{split}
\end{equation}
Since the $e/2$ quasihole/quasiparticle is just one more/less flux attached by the electrons and both of them are Abelian, we assume the  $e/2$ excitation in 331 state has the same properties as that in MR. This has been verified in the calculation of the topological spins as shown Fig.~\ref{figmpmr}.

For the $e/4$ excitation, the MR state and 331 state are distinct. Here we consider each of the edge has one $e/4$ quasi-hole. Then the root configuration is $\cdots  010101010 \cdots$. In this case, there are two topological sectors: $\bar{0}101|0101\bar{0}$ and $\bar{0}1010|101\bar{0}$. we have 
\begin{equation}
	\begin{split}
		&\bar{0}101|0101\bar{0}~~~h_\alpha=1/8 \\
		&\bar{0}1010|101\bar{0}~~~h_\alpha=1/8 \\ 
	\end{split}
\end{equation}

For quasi-hole MR state, since the total number of orbits is odd, the Fermi points are on top of the first orbit which is labelled by $\bar{0}$. It means only half of this orbit belongs to the subsystem.  For example, we consider the first sector of quasi-hole MR state, $M_A^0=-\frac{1+5}{2}=-3$, and $\bar{M_A}=\frac{1}{2}(-\frac{1+3+5+7/2}{2})=-\frac{25}{8}$, so $h_\alpha=-3+\frac{25}{8}=\frac{1}{8}$.  As a comparison, from the CFT description, the $e/4$ quasihole operator is expressed as $\sigma e^{i\sqrt{2}\phi/4}$ where $\sigma$ is the Majorana fermion field of Ising CFT and $\phi$ is the free chiral boson field. The $\sigma$ operator has conformal dimension $h = 1/16$ and thus the total dimension is $h_\alpha = 1/16 + \frac{(\sqrt{2}/4)^2}{2} = 1/8$.

For the $e/4$ excitation in 331 state, we obtain its conformal dimension from the CFT correlator. The ground state wave function could be written as~\cite{MR}
\begin{eqnarray}
\psi_{331}(\{z^{\uparrow}\},\{z^{\downarrow}\}) &=& \langle V^{+}(\{z_1^{\uparrow}\})\cdots V^{+}(\{z_N^{\uparrow}\})V^{-}(\{z_1^{\downarrow}\})\cdots V^{-}(\{z_N^{\downarrow}\})\rangle_{\text{spin}}
\left\langle \prod_{i=1}^{2N}e^{i\sqrt{\alpha}\phi_c(z_i)}\mathcal{O}_{bg}\right\rangle_{\text{charge}} \nonumber \\
&\simeq& \prod_{1 \leq i < j \leq N} (z_i^{\uparrow}-z_j^{\uparrow})^{\beta}
\prod_{1 \leq i < j \leq N} (z_i^{\downarrow}-z_j^{\downarrow})^{\beta}
\prod_{i,j}^N (z_i^{\uparrow}-z_j^{\downarrow})^{-\beta}
\prod_{1 \leq i < j \leq 2N} (z_i-z_j)^{\alpha} \nonumber \\
&\simeq& \prod_{1 \leq i < j \leq N} (z_i^{\uparrow}-z_j^{\uparrow})^{\alpha+\beta}
\prod_{1 \leq i < j \leq N} (z_i^{\downarrow}-z_j^{\downarrow})^{\alpha+\beta}
\prod_{i,j}^N (z_i^{\uparrow}-z_j^{\downarrow})^{\alpha-\beta}
\end{eqnarray}
where the spin vertex operators are $V^{\pm}(z) = e^{\pm i\sqrt{\beta}\phi_s(z)}$ and  $\mathcal{O}_{bg}$ is the background charge. Here the Gaussian factor has been neglected. For 331 state, $\alpha = 2, \beta = 1$ and thus the electron operator (carrying charge $e$ and spin 1/2) is 
\begin{eqnarray}
V^{\pm}(z) e^{i\sqrt{\alpha}\phi_c(z)} = e^{\pm i\phi_s(z)}e^{i\sqrt{2}\phi_c(z)}
\end{eqnarray}
The Abelian $e/4$ quasihole is written as 
$e^{\pm i\frac{1}{2}\phi_s(z)}e^{i\frac{\sqrt{2}}{4}\phi_c(z)}$ which has conformal dimension 
\begin{eqnarray}
h = \frac{1}{2} (\frac{1}{2})^2 + \frac{1}{2}(\frac{\sqrt{2}}{4})^2 = \frac{3}{16}.
\end{eqnarray}

The $e/4$ quasihole wave function (with a Lauglin quasihole in $\uparrow$ layer) can be written with chiral CFT correlator
\begin{eqnarray}
\psi_{331}(w,\{z^{\uparrow}\},\{z^{\downarrow}\}) &=& \left\langle 
e^{ i\frac{1}{2}\phi_s(w)}e^{i\frac{\sqrt{2}}{4}\phi_c(w)} V^{+}(\{z_1^{\uparrow}\})\cdots V^{+}(\{z_N^{\uparrow}\})V^{-}(\{z_1^{\downarrow}\})\cdots V^{-}(\{z_N^{\downarrow}\}) \prod_{i=1}^{2N}e^{i\sqrt{2}\phi_c(z_i)}\mathcal{O}_{bg}\right\rangle \nonumber \\
&\simeq& \prod_{i=1}^N (w - z_i^{\uparrow})^{1/2} \prod_{i=1}^N (w - z_i^{\downarrow})^{-1/2} \prod_{j=1}^{2N} (w - z_j)^{1/2}
\prod_{1 \leq i < j \leq N} (z_i^{\uparrow}-z_j^{\uparrow})^3
\prod_{1 \leq i < j \leq N} (z_i^{\downarrow}-z_j^{\downarrow})^3
\prod_{i,j=1}^N (z_i^{\uparrow}-z_j^{\downarrow}) \nonumber \\
&=& \prod_{i=1}^N (w - z_i^{\uparrow}) \prod_{1 \leq i < j \leq N} (z_i^{\uparrow}-z_j^{\uparrow})^3
\prod_{1 \leq i < j \leq N} (z_i^{\downarrow}-z_j^{\downarrow})^3
\prod_{i,j=1}^N (z_i^{\uparrow}-z_j^{\downarrow})
\end{eqnarray}
which demonstrates that a Laughlin quasihole in the upper layer has been created.

\section{$e/4$ Quasi-hole State on Cylinder}
\label{sec:app_qh}
 Because of the pairing nature of the Majorana mode, we can only create even number of $e/4$ quasiholes in the MR state. For MR state, creating one $e/4$ at $w$ means putting another one at infinity. Its wave function is
\begin{equation}
	\begin{split}
     |\psi^c_{5/2} \rangle &=\text{Pf}(\frac{\exp(\mathbf{\gamma z_i)}-\exp(\mathbf{\gamma w)}+\exp(\mathbf{\gamma z_j})-\exp(\mathbf{\gamma w)}}{\exp(\mathbf{\gamma z_i})-\exp(\mathbf{\gamma z_j})} ) \prod_{j<k} (\exp(\gamma \mathbf{z}_j)-\exp(\gamma \mathbf{z}_k))^2  e^{-\frac{1}{2}\sum_{i=1}^N x_i^2} e^{-\sum_i (N-1)\gamma \mathbf{z_i}}
    \end{split}
\end{equation}

 If we consider a pair $e/4$ quasiholes at $w_1$ and $w_2$, the wave function is
\begin{equation}
	\begin{split}
     |\psi^c_{5/2} \rangle &=\text{Pf}(\frac{(\exp(\gamma \mathbf{z_i})-\exp(\gamma \mathbf{w_1}))(\exp(\gamma \mathbf{z_j})-\exp(\gamma \mathbf{w_2}))+(\exp(\gamma \mathbf{z_i})-\exp(\gamma \mathbf{w_2}))(\exp(\gamma \mathbf{z_j})-\exp(\gamma \mathbf{w_1}))}{\exp(\mathbf{\gamma z_i)-\exp(\gamma z_j})} ) \\ & \prod_{j<k} (\exp(\gamma \mathbf{z}_j)-\exp(\gamma \mathbf{z}_k))^2  e^{-\frac{1}{2}\sum_{i=1}^N x_i^2} e^{-\sum_i (N-1)\gamma \mathbf{z_i}}
    \end{split}
\end{equation}  

For $e/4$ quasihole in the $\uparrow$ layer of the 331 state, the wave function is
\begin{equation}
\label{331eq1a}
	\begin{split}
		|\psi_{331} \rangle &=\prod_{i\in N_1} (\exp(\gamma \mathbf{z}_i)-\exp(\gamma \mathbf{w})) \prod_{i<j,i,j\in N_1} (\exp(\gamma \mathbf{z}_i)-\exp(\gamma \mathbf{z}_j))^3  \prod_{k<l,k,l\in N_2} (\exp(\gamma \mathbf{z}_k)-\exp(\gamma \mathbf{z}_l))^3  \\ 
	 & \prod_{i\in N_1,k\in N_2} (\exp(\gamma \mathbf{z}_i)-\exp(\gamma \mathbf{z}_k)) e^{-\frac{1}{2}\sum_{i=1}^N x_i^2} 
		  \cdot e^{-\sum_{i\in N_1} \frac{\gamma }{2}(4N_1-2)\mathbf{z_i}-\sum_{i\in N_2} \frac{\gamma }{2}(4N_2-2)\mathbf{z_i} } 
	\end{split}
\end{equation}
\end{widetext}

\end{document}